\documentclass[]{IEEEphot}

\usepackage{bm}
\usepackage{color}
\usepackage{algorithm}
\usepackage{algorithmic}
\usepackage{multirow}
\usepackage{threeparttable}
\usepackage{array}

\jvol{xx}
\jnum{xx}
\jmonth{June}
\pubyear{2009}

\newtheorem{definition}{Definition}

\begin{document}

\title{High-speed Privacy Amplification Scheme Using GMP in Quantum Key Distribution}

\author{B.~Z.~Yan, Q.~Li, H.~K.~Mao, X.~F.~Xue}

\affil{School of computer science and technology, Harbin Institute of Technology, Harbin 150006, China}  

\doiinfo{DOI: 10.1109/JPHOT.2009.XXXXXXX\\
1943-0655/\$25.00 \copyright 2009 IEEE}%

\maketitle

\markboth{IEEE Photonics Journal}{Volume Extreme Ultraviolet Holographic Imaging}



\begin{abstract}
Privacy amplification (PA) is the art of distilling a highly secret key from a partially secure string by public discussion. It is a vital procedure in quantum key distribution (QKD) to produce a theoretically unconditional secure key. The throughput of PA has become the bottleneck of most high-speed discrete variable QKD (DV-QKD) systems. Although some Toeplitz-hash PA schemes can meet the demand of throughput, their high throughput extremely depends on the high cost platform, such as MIC or GPU. From the comprehensive view of development cost, integration level and power consumption, CPU is a general low cost platform. However, the throughput of CPU based PA scheme is not satisfactory so far, mainly due to the conflict between the intrinsic serial characteristic of CPU and the parallel requirement of high throughput Toeplitz-hash PA scheme. In this paper, a high throughput modular arithmetic hash PA scheme using GNU multiple precision arithmetic library (GMP) based on CPU platform is proposed. The experimental results show that the throughput of our scheme is nearly an order of magnitude higher than the comparative scheme on the similar CPU platform, which is $135$ Mbps and $69$ Mbps at the block sizes of $10^6$ and $10^8$ on Intel i3-2120 CPU respectively. Moreover, our scheme can provide the best throughput in the field of DV-QKD, which is $260$ Mbps and $140$ Mbps at the block sizes of $10^6$ and $10^8$ on Intel i9-9900k CPU respectively.
\end{abstract}


\section{Introduction}

\begin{figure}[t]
	\centering
	\includegraphics[width=30pc]{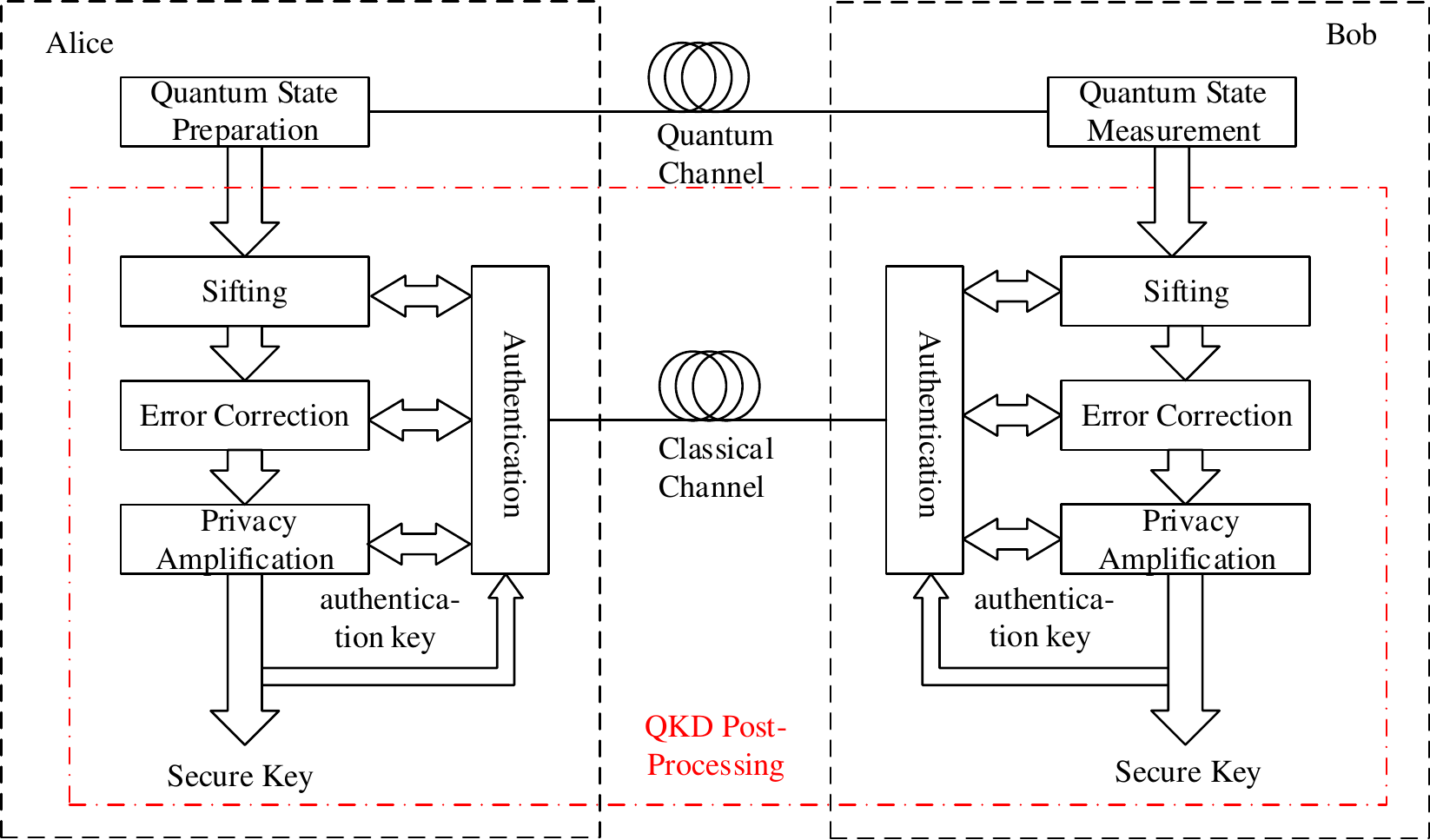}
	\caption{A general structure of the DV-QKD system}
	\label{fig_1}
\end{figure}

Quantum key distribution (QKD) is a notable technique which exploits the principles of quantum mechanics to perform the theoretically unconditional security key distribution between two remote parties, named Alice and Bob. The first practical QKD protocol is proposed by Bennet and Brassard in 1984 \cite{Bennett2014d}. Many protocols have been proposed since then, and these QKD protocols can be divided into discrete variable QKD (DV-QKD) and continuous variable QKD (CV-QKD) protocols \cite{Bru1998a,Bennett1992,Inoue2002,Zhang2019}. Comparing to CV-QKD, DV-QKD is applied much more widely and many high speed DV-QKD systems have been developed \cite{Wang2012,Yuan2018,Constantin2017c}. Therefore, we focus on the DV-QKD system in this paper. A general structure of the DV-QKD system is indicated as Fig.~\ref{fig_1}.

There are two major parts in a DV-QKD system: the quantum subsystem and the post-processing subsystem. The quantum subsystem is responsible of the preparation, transmission and measurement of quantum states. The post-processing subsystem distills Alice and Bob's fully uniform and secure key from their partially uniform and secure strings through a public but authentic channel. The post-processing subsystem mainly includes four parts: authentication~\cite{Li2016,Portmann2014}, sifting \cite{Li2015}, error reconciliation~\cite{Mao2019,Wang2018,Wang} and privacy amplification (PA). As one of the most important modules of the post-processing subsystem, the task of PA is to shrink the possible exposed information over the quantum channel and the classical channel to almost zero~\cite{Bennett1988a,Bennett1995}.

PA is the last link of producing the unconditional security key in QKD, which has great influence on the security of the final key. Bennet and Brassard proved that mapping a long corrected bit string to a much shorter final key via universal$_2$ hash function family can satisfy the security requirement of privacy amplification based on the classical information theory~\cite{Bennett1988a,Bennett1995}. While they only considered the situation that the adversary holds purely classical information about the raw key. Then Renner et al. proved that the same operation can ensure the security based on the quantum information theory, even if the eavesdropper can acquire and store the quantum information about the raw key instead of the classical information \cite{Renner2004}. Nevertheless, the security proof above is accomplished in the condition of asymptotic optimality, which means the block size of PA should be infinite. According to \cite{Tomamichel2012}, the block size of PA should be at least $10^6$, and the security of the final key will be enhanced as the increase of block size. The universal$_2$ hash function with such large block size leads to huge computation burden and results in the low throughput of PA. Therefore, PA has become the bottleneck of most high speed QKD systems. 

Universal$_2$ hash function is the kernel of PA, and the selection of universal$_2$ hash function family plays a decisive role in the computation burden \cite{Carter1979a,Mansour1993a}. Toeplitz hash function family has become the most popular choice in PA \cite{Krawczyk1994}. Many researchers focus on the optimization of Toeplitz-hash PA. J. Constantin et al. and S. S. Yang et al. respectively presented block parallel algorithm to accomplish Toeplitz-hash PA, and implemented it on field-programmable gate array (FPGA) \cite{Constantin2017c,Yang2017}. Except block parallel algorithm, fast Fourier transform (FFT) is an efficient algorithm for Toeplitz-hash PA due to the decrease of computational complexity, from ${\mathop{\rm O}\nolimits} (n^2)$ to ${\mathop{\rm O}\nolimits} (n\log n)$. B. Liu et al. implemented Toeplitz hash PA using FFT algorithm on a Many Integrated Core(MIC) platform \cite{Liu2016a}. Q. Li et al. implemented modified Toeplitz hash PA using FFT algorithm on a FPGA platform, reaching the processing speed of $116$ Mbps at the block size of $10^6$ \cite{Li2019}. R. Takahashi et al. utilized number theory transform (NTT) algorithm, a similar algorithm of FFT with the same computation complexity ${\mathop{\rm O}\nolimits} (n\log n)$ to implement Toeplitz hash PA on a MIC platform. This scheme achieved processing speed of $108$ Mbps at the block size of $10^8$ \cite{Yuan2018}. B. Y. Tang et al. implemented a large block size PA scheme with Toeplitz hash and FFT \cite{Tang2019}, and this scheme can reach $70$ Mbps throughput at $10^8$ block size using two Intel E5-2640 CPU (high-power server processor with stronger parallel performance than general-proposed CPU) and $128$ GB memory. X. Y. Wang et al. further improved the processing speed of Toeplitz hash PA to over $1$ Gbps with FFT on the graphics processing unit (GPU) for CV-QKD \cite{XiangyuWangYichenZhangSongYu2016a}. The schemes in \cite{Tang2019} and \cite{XiangyuWangYichenZhangSongYu2016a} can realize high throughput, but they both cost too much computing resources and their processing speed is influenced by the maximum compression ratio (the compression ratio refers to the ratio of the output and input key length of PA). Their maximum compression ratio will decrease as the processing speed increases, e.g., when the processing speed of the \cite{Tang2019} scheme reaches $70$ Mbps at $10^8$, the compression ratio of this scheme can not be larger than $0.125$. 

Summing up above schemes, some conclusions can be drawn: 1) Toeplitz hash function is the most popular method in PA and FFT/NTT algorithm can be used to speed up the Toeplitz-hash PA; 2) a variety of computing platforms have been utilized to design an efficient Toeplitz-hash PA scheme, and the platform with strong computing power can improve the throughput markedly. However, as the high performance platforms improve the throughput of the Toeplitz-hash PA, they also bring relatively large burden to the entire system and decrease the practicability of system. Comparing the characteristics of representative platforms for PA (in section 3.1), we can find that CPU is a general low cost platform considering development cost, integration and power consumption. However, the conflict between the intrinsic serial characteristic of CPU and the parallel requirement of high throughput Toeplitz-hash PA scheme leads to that the throughput of CPU based PA scheme is unsatisfactory so far.

Since Toeplitz-hash PA methods maybe insatiable for the low-cost CPU platform, more universal$_2$ hash families other than Toeplitz hash family should be studied and evaluated. Modular arithmetic hash is another kind of universal$_2$ hash family. It is baesd on modular arithmetic instead of matrix multiplication like Toeplitz hash family. The main operation of modular arithmetic hash, integer multiply operation, is the serial operation, so it is more suitable for the CPU platform. C. M. Zhang et al. once proposed an optimal multiplication algorithm for modular arithmetic hash PA, while the speed of this scheme is only $10.88$ Mbps at $10^7$, which is not good enough for most high speed DV-QKD systems \cite{Zhang2014}. Therefore, our research focus on the speed optimization of modular arithmetic hash function family, laying emphasis on the acceleration of large module multiplication in this hash computation. GNU multiple precision arithmetic library (GMP) is an efficient arithmetic library for general CPU architecture, which has effective combinatorial optimization on the speed of large module multiplication \cite{Granlund2015}. The speed optimization of large integer multiply and modular operation in our hash method using GMP is emphasisly studied. An efficient data pre-processing and post-processing method for arithmetic with GMP is also studied for the whole PA process acceleration. Then a modular arithmetic hash PA scheme is proposed using GMP in this paper. This scheme is designed based on the CPU platform and tested respectively on Intel i9-9900k and Intel i3-2120 platform. The experimental results on Intel i9-9900k reach the maximum throughput in DV-QKD as we know. It indicates the potential of modular arithmetic hash PA method instead of Toeplitz-hash PA method. More remarkably, our scheme on Intel i3-2120 improves the throughput of PA scheme on the similar platform (Intel i3-3220 in~\cite{Zhang2014}) by nearly an order of magnitude. With these experiment results, we compare the characteristics of existing PA schemes and analyze their applicable scope. Then, it can be seen that our scheme is the most practical PA scheme for most DV-QKD system at present.

The rest of this paper is organized as follows. Some related works are described and the reason of the huge computational cost of PA is explained in section 2 as the basis. In section 3, the characteristics of representative platforms are compared and the presented modular arithmetic hash PA scheme using GMP is described in details. In section 4, the experiment results and analysis are given. In section 5, some conclusions are drawn.

\section{Related Works}

\subsection{Universal Composable Security}

The universal composable security is a mathematical tool to describe the security of a complex system with multi cryptographic protocols. The final key generated by QKD is universal composable secure in the sense that this final key is distinguished with the ideal secret key except with a small probability. Thus, the QKD system with universal composable security can supply its secure key for arbitrary cryptographic equipments where the ideal secret key is expected. In this definition, the security of the final key generated by QKD is measured by the trace distance between the classical-quantum state with the real key and the classical-quantum state with the ideal key\cite{Renner2005}.
  
\begin{definition}[$\varepsilon$-secure key]
A secret key $K$ is said to be $\varepsilon$-secure with respect to an eavesdropper holding a quantum system $E$ if
	\begin{equation}
	\frac{1}{2}{\left\| {{\rho _{KE}} - {\rho _{UE}}} \right\|_1} \le \varepsilon,
	\end{equation}
where ${\rho _{KE}}$ represents the classical-quantum state with the final key $K$, and ${\rho _{UE}} = \sum\limits_{u \in U} {\frac{1}{{\left| U \right|}}\left| u \right\rangle \left\langle u \right| \otimes {\rho _E}} $ represents the classical-quantum state with the ideal key $U$. 
\end{definition}

\subsection{Privacy Amplification}

Privacy amplification in QKD aims to allow two parties, Alice and Bob, to distill a secure final key from a partially secure bit string. Privacy amplification based on quantum leftover hashing lemma can realize the universal composable security of the final key, and it is convenient for analyzing the influence of finite resource and the quantum state hold by the eavesdropper, named Eve, on the secure key in QKD. The inputs of two parties are identical random $n$-bit binary strings, $\mathbf{X}$ and $\mathbf{X'}$ generated by the error reconciliation module. Eve holds a quantum system $\bm{E}$ relative to $\mathbf{X}$. The definition of quantum leftover hashing lemma is given below. 

\begin{definition}[quantum leftover hashing lemma]
	
	Let ${\mathop{\rm g}\nolimits} $ be a compression function, which is randomly chosen from a universal$_2$ hash function family $\mathop{\rm G}: {\{ 0,1\} ^n} \to {\{ 0,1\} ^r}$, and $Y = {\mathop{\rm G}\nolimits} \left( X \right)$, then the following in-equation is valid for $\varepsilon  \ge 0$:
	\begin{equation}
	\frac{1}{2}{\left\| {{\rho _{YEG}} - {\rho _{{\rm{U}}EG}}} \right\|_1} \le \frac{1}{2} \times {2^{ - \frac{1}{2}(H_{\min }^\varepsilon ({\rho _{XE}}\left| E \right.) - r)}} + \varepsilon = \overline \varepsilon   ,
	\end{equation}
	where $H_{\min }^\varepsilon ( \cdot )$ is the $\varepsilon$-smooth min-entropy, and $H_{\min }^\varepsilon ({\rho _{XE}}\left| E \right.)$ means the bit number of uniform distribution key relative to Eve (with $\varepsilon$ failure probability) that can be extracted from the string $X$. 
\end{definition}

\begin{figure}[htbp]
	\centering
	\includegraphics[width=30pc]{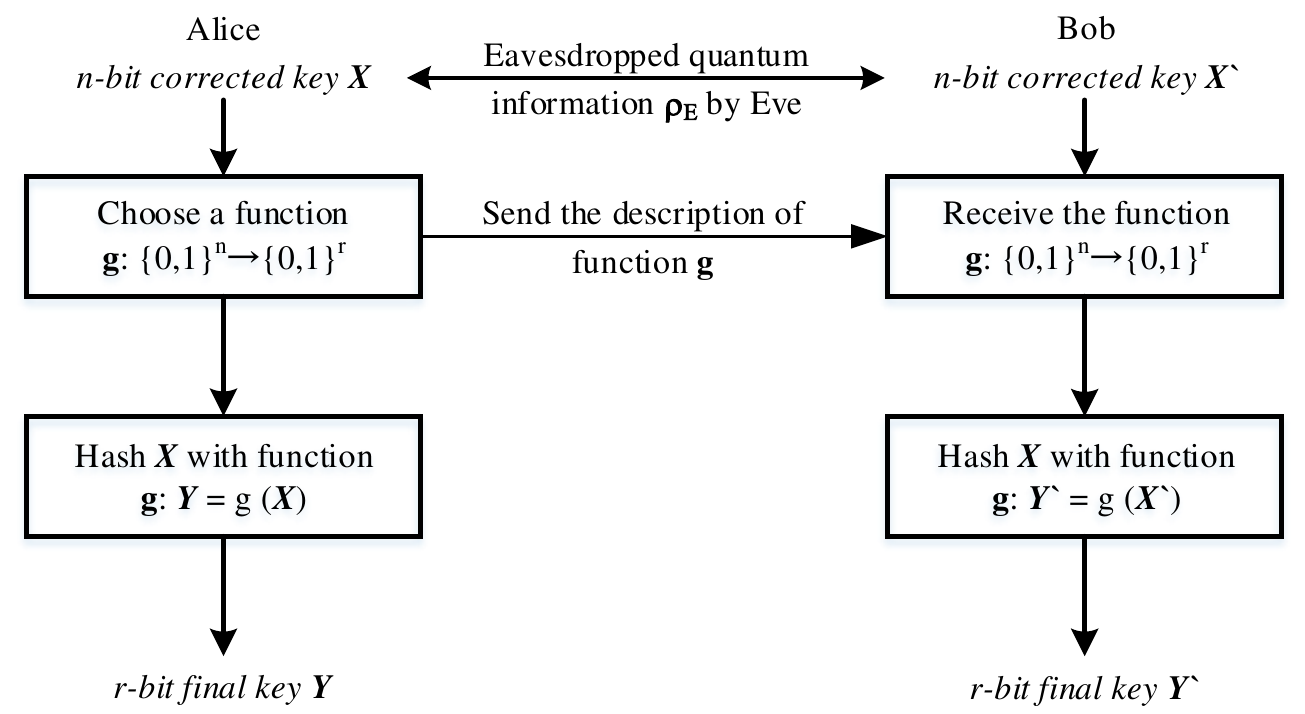}
	\caption{The procedure of privacy amplification in QKD}
	\label{fig_2}
\end{figure}

In PA based on quantum leftover hashing lemma, Alice and Bob wish to publicly choose a random compression function ${\mathop{\rm g}\nolimits} :{\{ 0,1\} ^n} \to {\{ 0,1\} ^r}$ from a universal$_2$ hash function family $\mathop{\rm G}$, then compress the string $\mathbf{X}$ and $\mathbf{X'}$ to generate the final key $\mathbf{Y}$ and $\mathbf{Y'}$ respectively, i.e., $\mathbf{Y} = {\mathop{\rm g}\nolimits} (\mathbf{X})$ and $\mathbf{Y'} = {\mathop{\rm g}\nolimits} (\mathbf{X'})$. If the length of final key $\mathbf{Y}$ satisfies the inequality relation in (\ref{eq1}), the final key $\mathbf{Y}$ is $ \varepsilon$-secure. Such procedure is indicated as Fig. \ref{fig_2}.

\begin{equation}
r \le H_{\min }^{\varepsilon /2}({\rho _{SE}}\left| E \right.) - 2{\log _2}\frac{1}{\varepsilon }.
\label{eq1}
\end{equation}

In the asymptotic sense ($n \to \infty $), the lower bound of $H_{\min }^\varepsilon ({\rho _{SE}}\left| E \right.)$ can be accurately estimated according to the protocol and system parameter. However, the block size $n$ is finite in the practical system, and the value of $H_{\min }^\varepsilon ({\rho _{SE}}\left| E \right.)$ can be only estimated at a worse lower bound. This leads to the decrease of the extreme length of final key. According to the conclusion of relevant study\cite{Tomamichel2012}, the block size $n$ should be at least $10^6$ to reduce this influence. The length of final key can be close to it in the asymptotic sense, when the block size $n$ reaches $10^8$.

The large block size leads to large computation requirement of universal$_2$ hash function. Therefore, the universal$_2$ hash function requiring less computation is extremely essential to implement high-speed PA.

\subsection{Universal$_2$ Hash Function}

Universal$_2$ hash function is the kernel of privacy amplification. The $\delta$-function is defined to evaluate universality of hash function.

\begin{definition}[$\delta$-function]
If $\mathop{\rm g}$ is a hash function from $\mathbf{A}$ to $\mathbf{B}$ and $x,y \in {\bf{A}}$, then
\begin{equation}
{\delta _{\mathop{\rm g}\nolimits} }(x,y) = \left\{ \begin{array}{l}
1{~~\rm{if}}~~x \ne y~{\rm{and}}~{\mathop{\rm g}\nolimits} (x) = {\mathop{\rm g}\nolimits} (y)\\
0{~~~~~~~\rm{otherwise}}
\end{array} \right..
\label{eq3}
\end{equation}

If element $x,y~{\rm{or}}~{\mathop{\rm g}\nolimits} $ is replaced by a set, i.e., $\rm{g}$ is replaced by a collection of hash functions $\mathbf{G}$ and $x$ is replaced by the set $\mathbf{A}$, then
\begin{equation}
{\delta _{\mathbf{G}}}({\bf{A}},y) = \sum\limits_{{\rm{g}} \in {\mathbf{G}}} {\sum\limits_{x \in {\bf{A}}} {{\delta _{\rm g}}(x,y)} }.
\label{eq4}
\end{equation}

\end{definition}

On this basis, the definition of universal$_2$ hash functions is given as follow.

\begin{definition}[universal$_2$ hash function]
Let ${\mathbf{G}}$ be a class of functions ${\mathbf{A}}$ from to ${\mathbf{B}}$. ${|\mathbf{A}|}$, ${|\mathbf{B}|}$ denotes the number of elements in ${\mathbf{A}}$, ${\mathbf{B}}$. The hash function family ${\mathbf{G}}$ is universal$_2$ if for any $x,y \in {\bf{A}}$ and $x \ne y$,

\begin{equation}
{\delta _{\bf{G}}}(x,y) \le {{\left| {\bf{G}} \right|} \mathord{\left/
		{\vphantom {{\left| {\bf{G}} \right|} {\left| {\bf{B}} \right|}}} \right.
		\kern-\nulldelimiterspace} {\left| {\bf{B}} \right|}}.
\label{eq5}
\end{equation}
 	
\end{definition}

$\mathbf{G}_{c,d}$ hash family defined as follows is an universal$_2$ hash family based on modular arithmetic.

\begin{definition}[$\mathbf{G}_{c,d}$ hash family]
	If the input size of $\mathbf{G}_{c,d}$ is $\left| {\bf{A}} \right| = {2^\alpha }$ , and the output size is $\left| {\bf{B}} \right| = {2^\beta }$ , then
	
	\begin{equation}
	{\bf{G}}_{c,d}^{} = \{ {{\mathop{\rm g}\nolimits} _{c,d}}:c,d \in {Z_{{2^\alpha }}},\gcd (c,2) = 1\} ,
	\label{eq6}
	\end{equation}
	
	\begin{equation}
	{{\mathop{\rm g}\nolimits} _{c,d}}(x) = {{\left\lfloor {(c \times x + d\bmod {2^\alpha })} \right\rfloor } \mathord{\left/
			{\vphantom {{\left\lfloor {(c \times x + d\bmod {2^\alpha })} \right\rfloor } {{2^{\alpha  - \beta }}}}} \right.
			\kern-\nulldelimiterspace} {{2^{\alpha  - \beta }}}}.
	\label{eq67}
	\end{equation}
	
\end{definition}

The multiplication of large numbers (over $10^6$ bits) is the most complex part of $\mathbf{G}_{c,d}$ function computation. Many optimization algorithms have been presented to accomplish the multiplication of large numbers, e.g., Karatsuba, Toom-Cook, Sch\"onhage and Strassen algorithms. These algorithms are suitable for different circumstances. GMP is an arithmetic library for multiple precision data, which can accomplish the multiplication of large numbers very fast via optimization of above algorithms. Therefore, we present a $\mathbf{G}_{c,d}$ hash privacy amplification scheme using GMP.

\section{Modular Arithmetic Hash PA Scheme Using GMP for CPU}

We first compare four representative platforms of PA in DV-QKD. The comparison shows that CPU is a comprehensively low cost platform for PA, while the performance of schemes on CPU is still unsatisfying. Then, A $\mathbf{G}_{c,d}$ hash PA scheme using GMP is designed for CPU platform. The principle and the advantage of this scheme for CPU are explained afterwards.
 
\subsection{The Comparison of Representative Platforms for PA}

The characteristics of representative platforms are summarized as development cost, integration level, power consumption listed in Table \ref{table_example}. 
\begin{table}[htbp]
	\renewcommand{\arraystretch}{1.3}
	\caption{The Characteristics of Representative Platforms of PA}
	\label{table_example}
	\centering
	\setlength{\tabcolsep}{5mm}{
	\begin{threeparttable}
	\begin{tabular}{ccccc}
		\hline
		\hline
		\bfseries Platform                    & \bfseries CPU        & \bfseries MIC & \bfseries GPU & \bfseries FPGA\\
		\hline
		\bfseries Development Cost \tnote{1}        & {\color{green} Good}  &  {\textcolor[rgb]{0.91,0.64,0.08} {Fair}} & {\textcolor[rgb]{0.91,0.64,0.08} {Fair}} & {\color{red} Poor} \\
		
		\bfseries Integration Level \tnote{2}    & {\textcolor[rgb]{0.91,0.64,0.08} {Fair}}   &	{\color{red} Poor} & {\color{red} Poor}  & {\color{green} Good} \\
		
		\bfseries Power Consumption     & {\color{green} Good}  & {\color{red} Poor} & {\color{red} Poor} & {\color{green} Good}  \\
		\hline
		\hline
	\end{tabular}
	\begin{tablenotes}
		\footnotesize
		\item[1] Development cost involves development difficulty, development cycle and device cost in QKD system design.  
		\item[2] Integration level means the number of control and calculation tasks in a QKD system that the platform can complete.  
	\end{tablenotes}
	\end{threeparttable}}
\end{table}

 \textbf{{Development cost}}: the development difficulty and development cycle of FPGA are highest and longest, because the basic operation and instruction of FPGA is too simple and even common algorithm implementations, e.g. FFT and large integer multiplication, on FPGA are always very complex. CPU and MIC are both based on x86 architecture and fully supported by various algorithm libraries, so the development cost is much lower. While the development of MIC needs clever parallel design for satisfactory performance and the device cost of MIC is much higher. So the development cost of MIC is higher than that of CPU. GPU also has its own special architecture. It also needs specialized design to make the most advantage of GPU parallel performance. 
 
 \textbf{Integration level}: the tasks of electronic device in a QKD system are system control and calculation. FPGA can complete both the system control and calculation tasks, so the integration level of FPGA is the highest. Then, CPU can only complete the calculation task but the system control. MIC and GPU are both designed for computing acceleration, so they always need a CPU to complete the calculation task together. Therefore, the integration level of MIC and GPU is the lowest.
 
  \textbf{Power consumption}: compared with CPU and FPGA, MIC and GPU contains a lot of computing cores and large amounts of memory to speed calculation. However, redundant structure and strong performance of MIC and GPU also bring much higher power consumption than CPU and FPGA.
 
 In summary, FPGA has the disadvantage of high development cost, and the block size of PA on FPGA is limited by $10^6$ is another important disadvantage. MIC and GPU both have disadvantages of poor integration level and high power consumption. Comparing to other platforms, CPU is a general low-cost platform without regard to its poor performance on PA. While the poor processing speed of PA scheme on CPU is due to that the common PA method (Toeplitz-hash PA) always improve its processing speed by parallel speedup. So the parallel platform, FPGA and MIC, performs better on these method. by contrast, our scheme is based on the large integer multiplication, which is more fit for the serial computing structure of CPU. The following subsection will introduce the principle of our scheme.     
 
\subsection{Modular Arithmetic Hash PA Scheme Using GMP}

\begin{figure}[htbp]
	\centering
	\includegraphics[width=17.5pc]{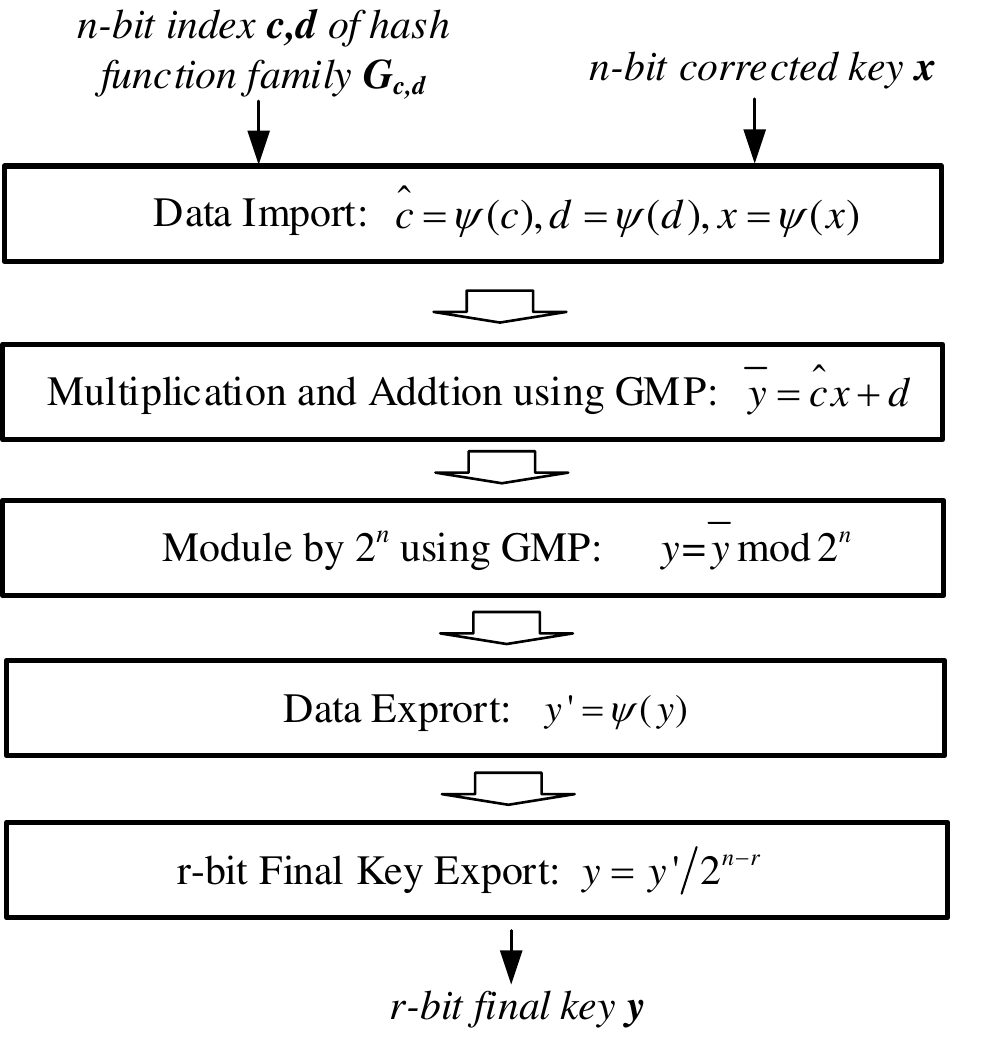}
	\caption{The procedure of the modular arithmetic $\mathbf{G}_{c,d}$ hash PA scheme using GMP}
	\label{fig_3}
\end{figure}

GMP is an arithmetic library written in C for arbitrary precision arithmetic on integers, rational numbers and floating points. In privacy amplification, the size of the operand is always over millions of bits. The advantage of GMP is the automatic selection of the suitable algorithm according to the operand size and the assembly code level optimization for the maximum operation speed. Therefore, a modular arithmetic $\mathbf{G}_{c,d}$ hash privacy amplification scheme using GMP is presented in this paper. The procedure of the scheme is indicated in Fig. \ref{fig_3}. 

\subsubsection{Data Import}

\begin{figure}[htbp]
	\centering
	\includegraphics[width=20pc]{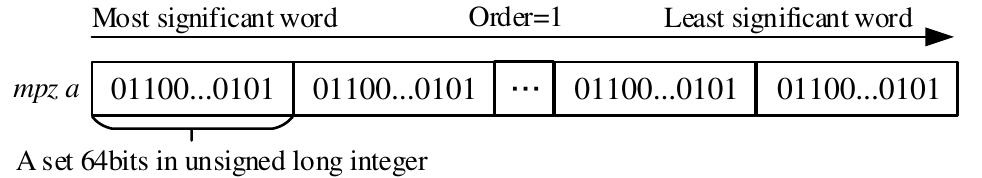}
	\caption{Data import procedure}
	\label{fig_4}
\end{figure}

The input data of privacy amplification is usually a bit string, while the operand of GMP multiplication is a single datum, whose type is \textit{mpz} instead of array or string. The format conversion process may be time-consuming in an inappropriate method. An efficient method for the format conversion is storing the bit string into an array of unsigned long integer data, and then importing it to a \textit{mpz} data using the function \textit{mpz\_import}. The data import procedure is indicated as Fig.  \ref{fig_4}.

The parameter \textit{order} is \textit{1} for most significant word first or \textit{-1} for least significant first.

\subsubsection{Multiplication and addition using GMP}

Multiplication is the most complex computation in $\mathbf{G}_{c,d}$ hash privacy amplification. The principles and advantages of GMP multiplication of large numbers are selectively analyzed.

\begin{figure}[htbp]
	\centering
	\includegraphics[width=15pc]{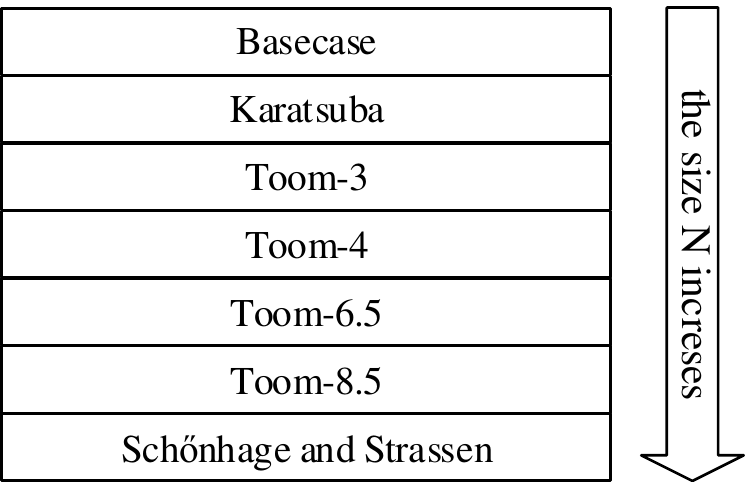}
	\caption{the algorithm selection based on the size N}
	\label{fig_5}
\end{figure}

Seven multiplication algorithms can be selected by GMP based on the multiplication size N. The relationship of the algorithm selection and the size N is indicated as Fig. \ref{fig_5}. 

As the multiplication size N in privacy amplification should be at least $10^6$, the multiplication algorithm for large size, i.e., Sch\"onhage and Strassen algorithm, is emphasisly analyzed.

The main principle of Sch\"onhage and Strassen algorithm is using FFT to transform the polynomial multiplication to the pointwise multiplication. NTT, FFT in number theory, is adopted in GMP to eliminate the influence of truncation error. The procedure of the algorithm is indicated as follow, which contains seven steps: fill zero, split, evaluate, pointwise multiply, interpolate, combine and carry bit. 

\begin{algorithm}[htb] 
\caption{Sch\"onhage and Strassen algorithm}
\label{SS_algorithm}
\begin{algorithmic}[1]	
\REQUIRE ~~
$a$ ($N$ bits) and $b$ ($N$ bits).
\ENSURE ~~
$c = a \times b$.
\STATE $a' = \phi \left( a \right),{\rm{ }}b' = \phi \left( b \right)$.
\STATE $\overline a  = \psi \left( {a'} \right),{\rm{ }}\overline b  = \psi \left( {b'} \right)$.
\STATE $\widehat a = NTT(\overline a ),{\rm{ }}\widehat b = NTT(\overline b )$.
\STATE $\widehat c = \widehat a \cdot \widehat b$.
\STATE $\overline c  = INTT\left( {\widehat c} \right)$.
\STATE $c' = {\psi ^{ - 1}}(\overline {c)} $.
\STATE $c = \gamma \left( {c'} \right)$.
\end{algorithmic}	
\end{algorithm}

This algorithm is explained progressively:1) fill zero: padding $a$ and $b$ with high zero to the length $2^N$ bits. 2) split: splitting $a'$ and $b'$ into $2^k$ pieces of $M = {{2N} \mathord{\left/{\vphantom {{2N} {{2^k}}}} \right.\kern-\nulldelimiterspace} {{2^k}}}$ bits each. 3) evaluation: calculating $2^k$ points NTT of $\overline a$ and $\overline b$. 4) point-wise multiply: multiplying $\widehat a$ and $\widehat b$ point-wisely, the length of each multiply should be $N' = 2M + k + 3$ at least. 5) interpolate: calculating $2^k$ points NTT of $\widehat c$. 6) combine: combining the $2^k$ pieces of $\overline c $. 7) carry bits: carry operation according to the size of the base. It is worth noting that, in step 4, the length $N'$ of multiplication may be still very large, so GMP multiplication will be used again. Then Sch\"onhage and Strassen, Toom-Cook or other algorithms will be selected based on the length $N'$.

In the design of implementation scheme, the above algorithm can be used by the function named \textit{mpz\_mul}. While the function \textit{mpz\_addmul} is adopted in this scheme, because it can accomplish one addition and one multiplication at one time. It is more efficient to implement $y = cx + d$ in this PA scheme.

\subsubsection{Module by $2^n$}

In this PA scheme, a modular arithmetic by $2^{n}$ is necessary. Other than a normal modular arithmetic needs division operation, a $2^{n}$ modular arithmetic can be accomplished only by bitwise operation. Therefore, a specialized function named \textit{mpz\_mod\_2exp} is adopted instead of \textit{mpz\_mod} to accelarate the whole process. 

\subsubsection{Data Export}

As similar with Data Import, data export is to transform the large integer result in type \textit{mpz} to an array of the type unsigned long using the function \textit{mpz\_export}.

\subsubsection{Export r bits final key}

A transformation from the unsigned long integer array to a bit string is implemented in this step, and only the most significant r bits of the bit string can be exported. The value of $r$ is calculated by the conclusion of the GLLP security theory presented by D.  Gottesman et al. based on the practical system parameter \cite{Gottesman2002}.

\section{Results and Analysis}

The $\mathbf{G}_{c,d}$ hash PA scheme above is implemented on two different CPU platforms. The throughput results of our scheme on two platforms are given in TABLE \ref{table_2}.

\begin{table}[htbp]
	\renewcommand{\arraystretch}{1.3}
	\caption{the throughput result of $\mathbf{G}_{c,d}$ hash PA scheme}
	\label{table_2}
	\centering
	\begin{tabular}{ccccc}
		\hline
		\hline
		\multirow{2}{*}{\bfseries Blocksize(bits)} & \multicolumn{2}{c}{Scheme on Intel i9-9900k CPU} & \multicolumn{2}{c}{Scheme on Intel i3-2120 CPU}\\
		\cline{2-5}
		& Time(ms) & Throughput(Mbps) & Time(ms) & Throughput(Mbps) \\
		\hline
		\bfseries 1M & {3.81} & {262.13} & {7.38} & {135.56} \\
		
		\bfseries 2M & {8.19} & {244.15} & {16.35} & {122.31} \\
		
		\bfseries 4M & {18.23} & {219.39} & {37.35} & {107.09} \\
		
		\bfseries 8M & {40.75} & {196.33} & {87.71} & {91.21} \\
		
		\bfseries 10M & {53.23} & {187.86} & {112.66} & {88.76} \\
		
		\bfseries 16M & {89.99} & {177.8} & {186.15} & {85.95} \\
		
		\bfseries 32M & {202.61} & {157.94} & {435.02} & {73.56} \\
		
		\bfseries 64M & {427.92} & {149.56} & {879.97} & {72.73} \\
		
		\bfseries 100M & {709.42} & {140.96} & {1437.40} & {69.57} \\
		\hline
	\end{tabular}
\end{table}

As we can see from the results: 

1) Our scheme can cover the block size from $10^6$ to $10^8$. This is the first time that the PA scheme on general-proposed CPU platform reaches the block size of $10^8$ . 

2) Our scheme on Intel i9-9900k CPU gets the best throughput result in DV-QKD, which can be testified by the following comparison. 

3) Our scheme on Intel i3-2120 CPU gets less but still sufficient throughput result.

Then, we will compare our schemes with the best scheme designed on representative platforms. The advantages of our scheme in practical DV-QKD system can be illustrated by following comparisons and analysis.

\subsection{The comparison between our scheme and others}

\begin{figure}[htbp]
	\centering
	\includegraphics[width=30pc]{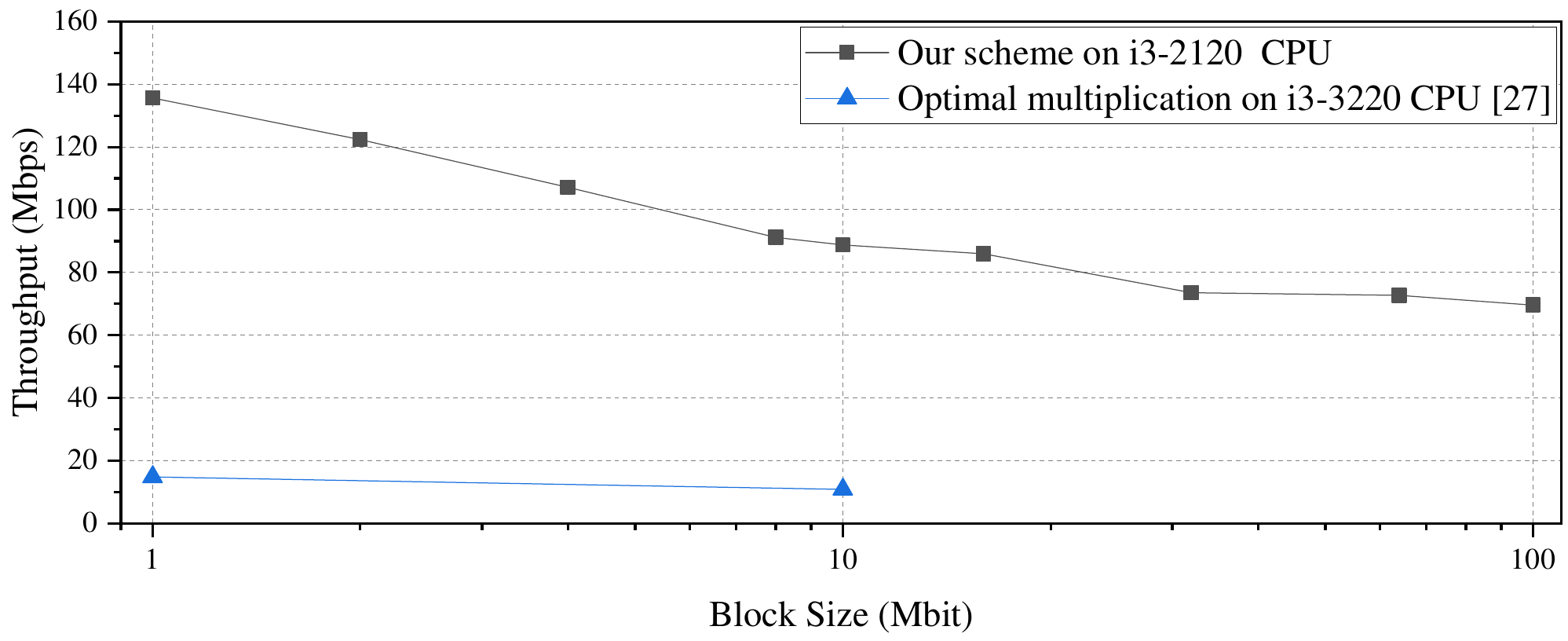}
	\caption{the throughput comparison between our scheme and the best PA scheme on the similar CPU platform}
	\label{fig_8}
\end{figure}

First, we will compare the $\mathbf{G}_{c,d}$ hash PA scheme with other scheme on the similar platform. Though the costs of our scheme on i3-2120 and the optimal multiplication scheme are similar as shown in TABLE \ref{table_3}. While The performance of our scheme is much better as indicated in Fig. \ref{fig_8}. The throughput of our scheme can reach ten times of the optimal multiplication scheme at the same block size. 

\begin{table}[htbp]
	\renewcommand{\arraystretch}{1.3}
	\newcommand{\tabincell}[2]{\begin{tabular}{@{}#1@{}}#2\end{tabular}}
	\caption{Key parameters of the compared schemes}
	\label{table_3}
	\centering
	\begin{tabular}{ccccccc}
		\hline
		\hline
		PA scheme & Hash & Based Method & Platform & \tabincell{c}{Number of\\ Cores} &\tabincell{c}{ Basic\\  Frequency} & Memory \\
		\hline
		\tabincell{c}{Our scheme\\ on i9-9900} & $\mathbf{G}_{c,d}$ & GMP multiply & Intel i9-9900k & 8 & 3.60 GHz & 16 GB \\ 
		
		\tabincell{c}{Our scheme\\ on i3-2120} & $\mathbf{G}_{c,d}$ & GMP multiply & Intel i3-2120 & 2 & 3.30 GHz & 8 GB \\
		
		\tabincell{c}{Toeplitz NTT\\ on MIC \cite{Yuan2018}}   & Toeplitz & NTT & \tabincell{c}{Intel Xeon E5-2620v2\\+ Intel Xeon Phi 7120A} & \tabincell{c}{2\\+ 61}& \tabincell{c}{2.10GHz\\+ 1.24GHz} &  \tabincell{c}{16GB\\+ 128GB} \\
		
		\tabincell{c}{Optimal multiplication\\ scheme \cite{Zhang2014}} & $\mathbf{G}_{c,d}$ & Optimal multiply & Intel i3-3220 & 2 & 3.30 GHz & 4GB \\
		\hline 
		\hline
	\end{tabular}
\end{table}

\begin{figure}[htbp]
	\centering
	\includegraphics[width=30pc]{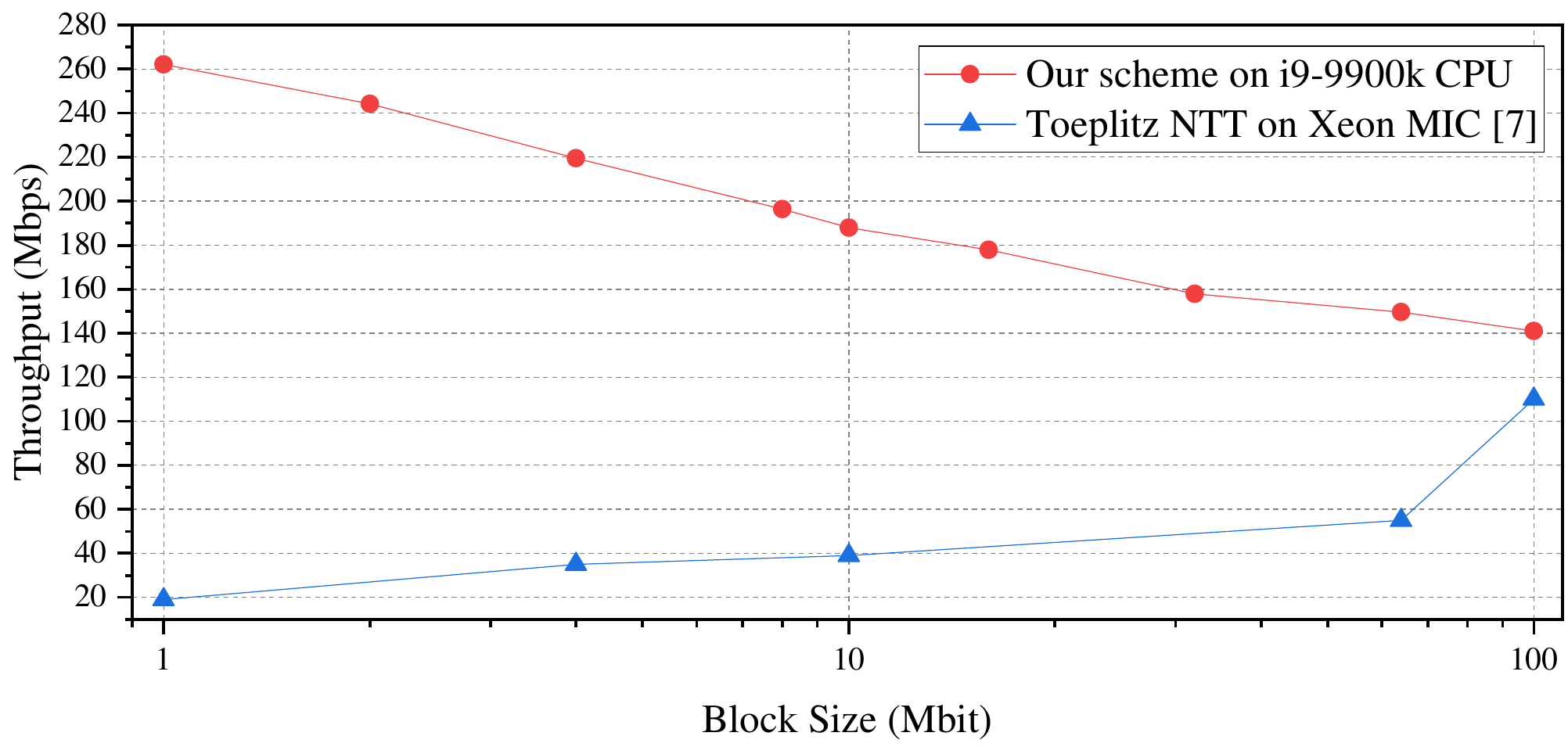}
	\caption{the throughput comparison between our scheme and the best PA scheme on MIC}
	\label{fig_7}
\end{figure}

Then, we compare our scheme with the best PA scheme on MIC. Though the cost of the MIC scheme is more as shown in TABLE \ref{table_3}. The performance comparison with our scheme is indicated as Fig. \ref{fig_7}. The throughput of the NTT Toeplitz scheme improves as the block size increases, that is benefit by the parallel advantage of MIC scheme. But this tendency can not continue because of the limitation of read/write speed of memory. Considering the total range of common block size, the performance of our scheme is better than the NTT Toeplitz scheme with much less cost.

\begin{figure}[htbp]
	\centering
	\includegraphics[width=30pc]{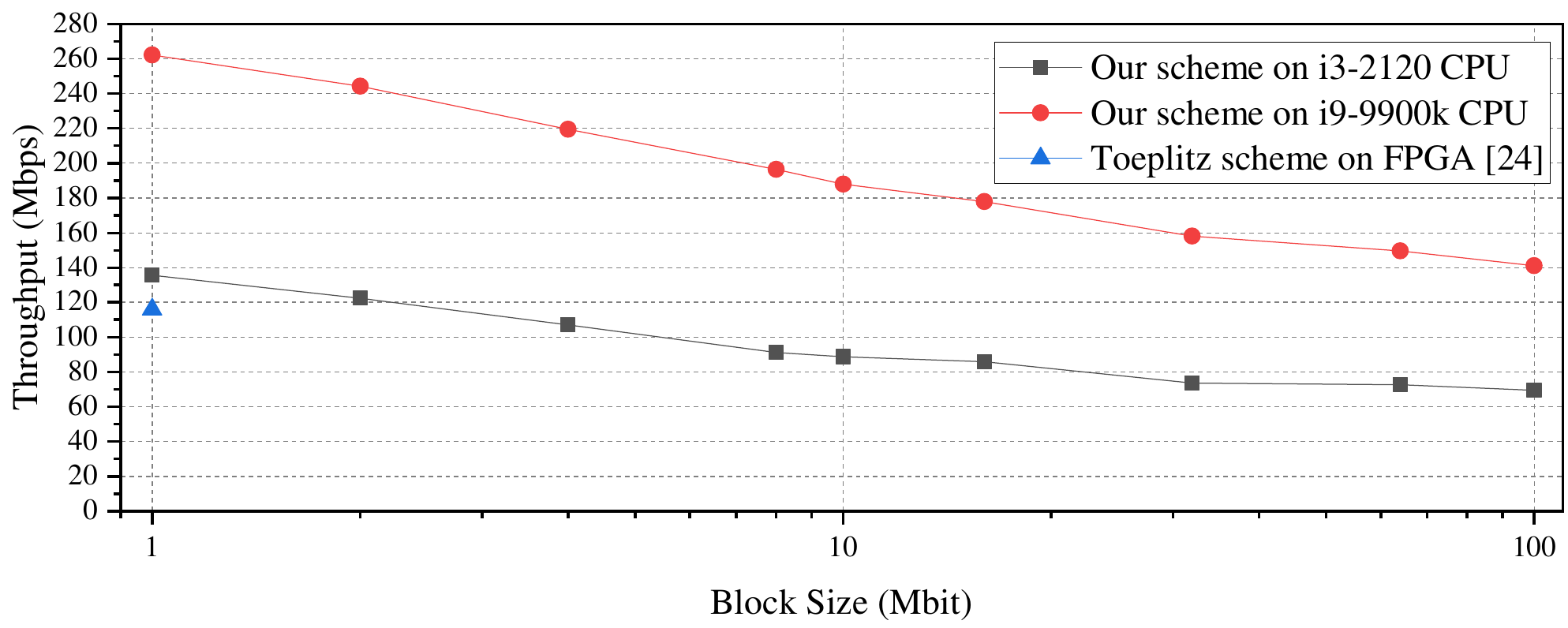}
	\caption{the throughput comparison between our scheme and the best PA scheme on FPGA}
	\label{fig_6}
\end{figure}

We last compare our scheme with the best PA scheme on FPGA. As mentioned earlier, the scheme on FPGA is limited by the block size. The throughput comparison is indicated as Fig. \ref{fig_6}.

Beyond the scheme compared above, the schemes in \cite{Tang2019} and \cite{XiangyuWangYichenZhangSongYu2016a} also have satisfactory throughput and block size. While these schemes  both sacrifice the compression ratio to improve the performance. Their maximum compression ratio, which is usually lower than $20\%$, may not satisfy the need of the evolving QKD system.

Through above comparison, these conclusion can be drawn: 1) Our scheme obviously improves the block size of PA, and the throughput can be improved by nearly an order of magnitude on the similar CPU platform. 2) The results on the Intel i9-9900k has reach the best throughput in the DV-QKD.

\section{Conclusions}
The main issue of exciting PA schemes in QKD is analyzed in this paper. Targeting at the problem that Toeplitz hash PA schemes are unable to satisfy the demand of the developing QKD system, we focus on modular arithmetic hash function to design a PA scheme. Then the characteristics of representative platform for PA are compared and CPU is selected to be a suitable platform for practical QKD systems. Then a modular arithmetic based $\mathbf{G}_{c,d}$ hash privacy amplification scheme using GMP is presented. 

The major problem of PA in QKD is analyzed in this paper, that is too huge amount of PA computation due to large block size. This makes the throughput of PA become a bottleneck of QKD system. various computing platforms have been used to optimize the performance of PA. This paper compares the characteristics of representative platforms for PA and points that some platforms, e.g. GPU and MIC, add significantly to the cost and reduce the practicability of QKD system. In contrast, CPU is a general low-cost platform for QKD system. However, the throughput of CPU based PA scheme is not satisfactory so far, mainly due to the conflict between the intrinsic serial characteristic of CPU and the parallel requirement of high throughput Toeplitz-hash PA scheme. 

Then a modular arithmetic based $\mathbf{G}_{c,d}$ hash privacy amplification scheme using GMP is presented for CPU platform. This scheme is implemented and tested on two different CPU platforms. The results indicate that this scheme 1) improves the block size of PA and the throughput by nearly an order of magnitude on the similar CPU platform; 2) reaches the best throughput result of existing PA schemes in the DV-QKD systems; 3) verifies the high efficiency of modular arithmetic hash PA method and using GMP in $\mathbf{G}_{c,d}$ modular arithmetic hash; 4) is the most practical PA scheme for DV-QKD considering performance, development cost, integration level and power consumption.

\section*{Acknowledgements}
The authors wish to thank the anonymous reviewers for their valuable suggestions.  


\bibliographystyle{IEEEtran}
\bibliography{IEEEabrv,QKD}



\end{document}